\documentclass[runningheads]{llncs}
\usepackage{hyperref}
\usepackage{graphicx}
\usepackage{graphicx}

\usepackage{amsmath}
\usepackage{array}
\usepackage{caption}
\usepackage{floatrow}   
\usepackage{subcaption}
\usepackage{graphicx,xcolor} 
\usepackage{breqn}
\usepackage[framemethod=tikz]{mdframed}
\usepackage[tableposition=top]{caption}
\usepackage{float}
\floatstyle{plaintop}
\restylefloat{table}
\usepackage{arydshln}

\hypersetup{draft}

\usepackage{cite}

\begin{document}

\title{Semi-Supervised Training of Optical Flow Convolutional Neural Networks
	in Ultrasound Elastography\thanks{Supported by NSERC Discovery Grant RGPIN 04136.}}
\titlerunning{Semi-Supervised Training of Optical Flow CNNs in USE}
	%
	\author{Ali K. Z. Tehrani \and
		Morteza Mirzaei \and
		Hassan Rivaz}
	\authorrunning{Ali K. Z. Tehrani}
	%
	\institute{Department
		of Electrical and Computer Engineering, Concordia University, Canada}
	\maketitle              
	\begin{abstract}
		Convolutional Neural Networks (CNN) have been found to have great potential in optical flow problems thanks to an abundance of data available for training a deep network. The displacement estimation step in UltraSound Elastography (USE) can be viewed as an optical flow problem. Despite the high performance of CNNs in optical flow, they have been rarely used for USE due to unique challenges that both input and output of USE networks impose. Ultrasound data has much higher high-frequency content compared to natural images. The outputs are also drastically different, where displacement values in USE are often smooth without sharp motions or discontinuities. The general trend is currently to use pre-trained networks and fine-tune them on a small simulation ultrasound database. However, realistic ultrasound simulation is computationally expensive. Also, the simulation techniques do not model complex motions, nonlinear and frequency-dependent acoustics, and many sources of artifact in ultrasound imaging. Herein, we propose an unsupervised fine-tuning technique which enables us to employ a large unlabeled dataset for fine-tuning of a CNN optical flow network. We show that the proposed unsupervised fine-tuning method substantially improves the performance of the network and reduces the artifacts generated by networks trained on computer vision databases.   
		
	\end{abstract}
	\section{Introduction}
	Ultrasound is one of the most widely used modality in medical imaging, and is the preferred modality in image-guided interventions \cite{azizi2018learning,zhou2019handbook,zhuang2019region}. UltraSound Elastography (USE) is an imaging technique which provides relative stiffness properties of the tissue, and as such, provides additional guidance during interventions. Free-hand palpation is one of the most popular methods in USE due to simplicity and availability. The basic idea of free-hand palpation is that the operator compresses the tissue by ultrasound probe, and the images before and after the compression are compared to obtain the displacement map \cite{ophir1999elastography}. Due to the fact that most compression is in the axial direction, axial displacement contains more available information than the lateral one. The axial displacement map is used to obtain the strain map, which is generally inversely proportional to the elastic modulus.

	Convolutional Neural Networks (CNN) have been proven useful in optical flow estimation. Many network architectures such as FlowNet \cite{dosovitskiy2015flownet}, FlowNet2 \cite{ilg2017flownet}, PWC-Net \cite{sun2018pwc} and LiteFlowNet \cite{hui2018liteflownet} have been proposed. The displacement estimation step of the USE can be performed using optical flow CNNs \cite{kibria2018gluenet,peng2018convolution,tehrani2020displacement,wu2018direct,peng2020neural,gao2019learning}. However, computer vision images and ultrasound data are generally different in characteristics and the objectives. Computer vision images may contain small objects with a very different optical flow from the background (for example: a hand moves and the background is fixed). Whereas in USE, the movement is generally smooth and continuous. Another difference lies in the objective of the two tasks. The objective is to find sharp and accurate optical flows in computer vision, whereas in USE, the main goal is to obtain a differentiable displacement field. These differences led to the fact that the strain map generated by optical flow CNNs trained on computer vision images have lower bias but with higher variance compared to traditional elastography algorithms \cite{tehrani2020displacement}. The lower bias of CNNs results in high contrast images but the high variance is amplified in the spatial differentiation step. Fine-tuning is a viable options to improve the network performance and reduce this variance \cite{kibria2018gluenet,peng2018convolution,tehrani2020displacement,peng2020neural}.     
	
	Many researchers have tried to adopt optical flow CNNs for USE using supervised fine-tuning. The general trend among researchers is to use pre-trained networks and fine-tune them with generated simulation datasets with known ground truth \cite{peng2018convolution,tehrani2020displacement,peng2020neural,evain2020pilot}. They used pre-trained well-known optical flow CNNs such as FlowNet2, PWC-Net and LiteFlowNet, and fine-tuned them using supervised techniques. The structure of the networks are also modified to address the differences of computer vision and USE in the inputs \cite{tehrani2020displacement}.  
	
Unsupervised fine-tuning is a more appropriate option for several reasons. First, simulation techniques entail several finite element and interpolation steps, which render the accuracy of sub-pixel ground truth displacement field inaccurate. Second, the simulation database often cannot model non-linear deformation and acoustic behaviors. Last but not least, the fine-tuning may cause forgetting effects \cite{li2017learning}, if the imaging parameters of ultrasound device is not close to the simulation data to the point that the fine-tuning with simulation deteriorates results on real data \cite{tehrani2020displacement}. By using unsupervised techniques, the network can be fine-tuned to any target domain, i.e. different ultrasound machines and different organs. 
	
	In this paper, we propose a novel unsupervised technique to fine-tune pre-trained optical flow CNNs using real ultrasound images. Our method can be considered as a form of semi-supervised learning since the pre-trained network is trained by labeled data, whereas the fine-tuning is done using data with unknown ground truth. 
	We use LiteFlowNet \cite{hui2018liteflownet} since it is light and has shown good performance in optical flow. However, the proposed framework can be applied to any optical flow CNN. The network estimates 2D displacements for strain values ranging from 0.5 to 5 \%, the performance of the algorithm in transient elastography \cite{sandrin2003transient} where displacements are very small is an area of future work. Our contribution can be summarized as follows:

	\begin{enumerate}
			\item Our results show that training on computer vision images is not enough since the statistics of ultrasound RF data and physics of the displacement field are different in these two domains.
		\item We propose an unsupervised fine-tuning method in elastography.
		\item We use real ultrasound images for fine-tuning, thanks to our unsupervised technique which does not need ground truth displacements. 
		\item We propose an automatic frame and region selection algorithm which enables the user to employ real ultrasound images without any supervision and expertise. 
		\item We propose a novel loss function, considering statistics of RF data and physics of the displacement field.
	\end{enumerate}

	\section{Material and Methods}

	\subsection{Unsupervised training of optical flow networks}

	A critical component of unsupervised techniques is the loss function, which can be expressed as \cite{godard2017unsupervised,meister2018unflow,ren2020unsupervised,wang2018occlusion}:

	\begin{equation}
	Loss = loss_{d}+loss_{s}+loss_c
	\end{equation}
	
	where $loss_{d}$ is the data loss, $loss_{s}$ is the smoothness loss which can also be described as smoothness regularization, and $loss_c$ is the consistency loss,  which shows how different the forward and backward flows are. Data loss can be described as the difference between the first image and the warped second image. Smoothness loss controls smoothness of the displacement and usually first- and second-order derivatives are utilized for this loss \cite{meister2018unflow,ren2020unsupervised,wang2018occlusion}. In \cite{godard2017unsupervised}, a combination of $L{1}$ norm and structural similarity ($SSIM$) is employed for data loss, an edge-aware smoothness regularizer is utilized as the smoothness loss, and $L_1$ norm is used for consistency loss. In \cite{meister2018unflow}, the robust generalized Charbonnier penalty \cite{sun2014quantitative} of census transform \cite{zabih1994non} is employed as the data loss. The Charbonnier penalty of forward and backward displacement as consistency loss have also been exploited in \cite{meister2018unflow}. Finally, this paper also used an occlusion mask to remove the occluded pixels from the loss terms to avoid back propagation of occluded regions. 

	\subsection{Proposed Method}

	Inspired by other unsupervised optical flow networks, we propose a fine-tuning strategy well-adapted to USE. Let $I_1$ and $I_2$ be the first and the second images, $w_f$ and $w_b$ be forward ($I_1$$\rightarrow$$I_2$) and backward flows ($I_2$$\rightarrow$$I_1$). We define the data loss as:
	\begin{equation}
	loss_d=\left \langle \Phi(I_1-\widetilde{I_2}) \right \rangle_{O_f}         
	\end{equation}
	where $\left \langle . \right \rangle_{O_f}$ is the mean of non-outlier pixels.  $\widetilde{I_2}$ is the warped $I_2$ toward $I_1$ using $w_f$ and $\Phi$ denotes Charbonnier penalty:
	\begin{equation}
	\Phi(x)=(x^2+\varepsilon)^\gamma    
	\end{equation} 
	We set $\gamma$ to 0.2 similar to the fine-tuning loss in \cite{ilg2017flownet,sun2018pwc,tehrani2020displacement} and $\varepsilon$ denotes a small number. There is no occlusion in USE. We borrow ideas from occlusion detection to find the outlier regions of displacement estimates and exclude them in all loss terms. In order to find outlier displacement estimates, we compare forward and backward displacement using the following equation:
	\begin{equation}
	O_f=\left | w^f+w^b \right |<\alpha 
	\end{equation}
	we set $\alpha$ empirically to 1 as we observed that in case of outlier the difference between forward and negative of the backward displacement is much larger than 1. The $O_f$ image can be considered as a mask that selects reliable regions for the loss function. In unsupervised training, frame selection is critical since many pairs of frames are not suitable for displacement estimation due to a very large decorrelation between their ultrasound data, and grossly incorrect displacement estimates can back propagate wrong values to the network. In order to do frame selection during the training, image pairs wherein the $O_f$ mask is 0 in more than 50\% of pixels are excluded. The outlier mask can be considered as a hard threshold consistency loss.

\begin{figure}[!t]
		\centering
		\includegraphics[height=0.22\textheight]{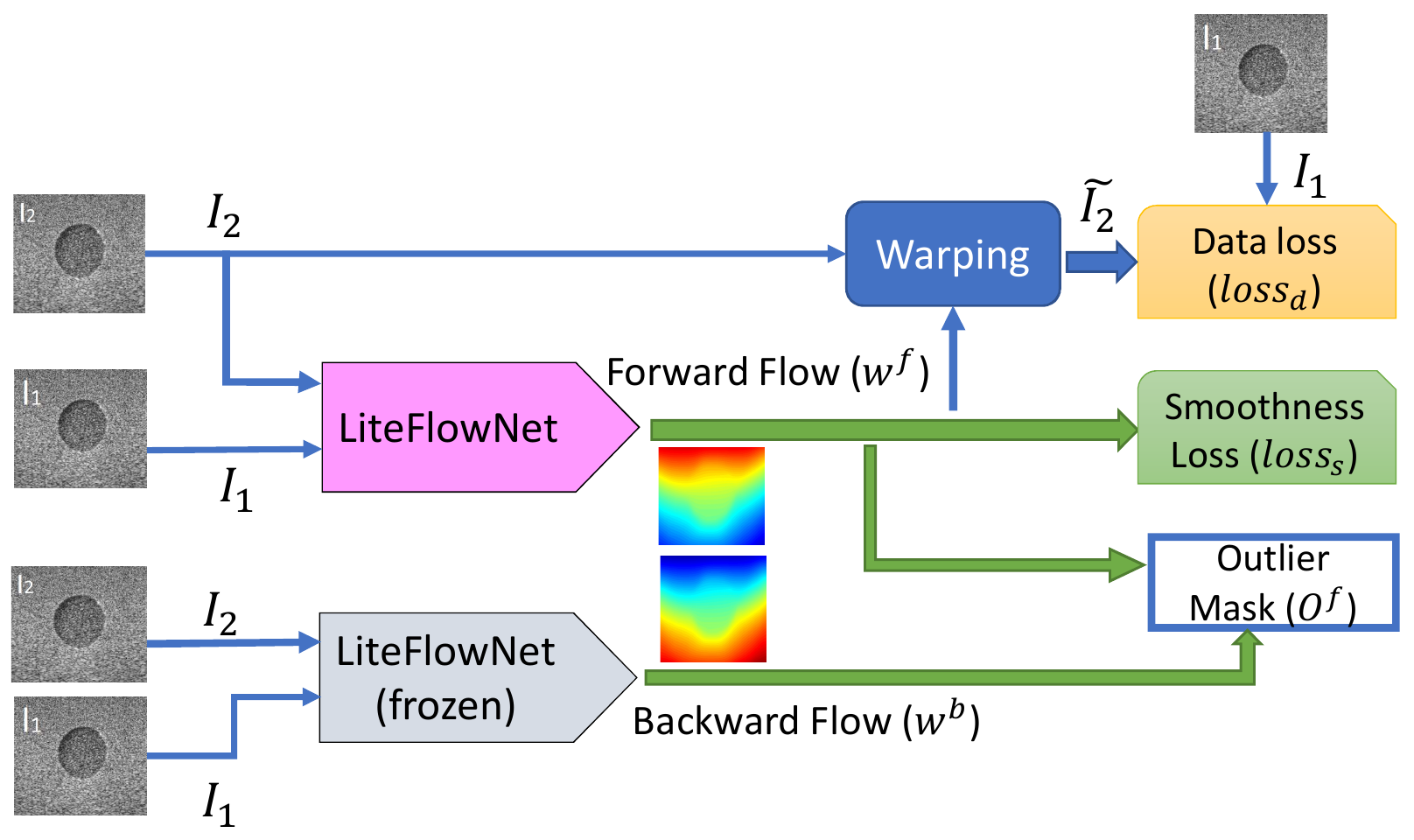}
		\centering
		\caption{Proposed network structure for unsupervised training. }
		\label{fig:Unlight}

	\end{figure}

	Regarding the smoothness loss,  the derivative of axial displacement is often the main concern. This derivative operation amplifies variance of the displacement estimates, reducing the contrast to noise ratio (CNR). In optimization-based elastography methods, smoothness constrains are imposed on the axial displacement \cite{hashemi2017global,mirzaei2019combining}. Here, we enforce smoothness on both displacement and its derivative. The latter is insensitive to affine deformations and performs better in the boundaries \cite{meister2018unflow}. Let axial displacement be $w^f_a$, and $a$ and $l$ denote axial and lateral direction, respectively. The first order smoothing loss can be given as:
	\begin{equation}
	loss_s^1=\lambda _1\left \langle\Phi\left \{\frac{\partial }{\partial a}w^f_a-<\frac{\partial }{\partial a}w^f_a>\right \}\right \rangle_{O_f}+\lambda _2\left \langle\Phi\left \{\frac{\partial }{\partial l}w^f_a-<\frac{\partial }{\partial l}w^f_a>\right \}\right \rangle_{O_f}    
	\end{equation}  
where $\lambda _1$ and $\lambda _2$ are weights associated to the axial and lateral derivative, respectively and $\left \langle.\right \rangle$ denotes mean. The average of derivatives are subtracted from the regularization term to reduce the regularization bias \cite{mirzaei2019combining,tehrani2020displacement}. 
	We also consider to penalize changes in the second order axial derivative of axial displacement:
	\begin{equation}
	loss_s^2=\lambda _3\left \langle\Phi\left \{\frac{\partial^2 }{\partial a^2}w^f_a\right \}\right \rangle_{O_f}    
	\end{equation}
The network structure for unsupervised training is shown in Fig. \ref{fig:Unlight}. The final loss function for training can be written as:
	\begin{equation}
	Loss=loss_d+ loss_s^1+loss_s^2
	\end{equation}
It should be noted that LiteFlowNet is a multi-scale network with
intermediate outputs. For training, intermediate loss and labels are
required, but as suggested by \cite{tehrani2020displacement}, we only consider the last output since only small changes are required in fine-tuning.
 
	\subsection{Training and Practical Considerations}
	
	Unlike computer vision images, a large input size (for example, 1920$\times$768 in this work) is required in USE to maintain the high frequency information of the radio frequency (RF) data. This is a limiting factor for current commercial GPUs, which generally have less than 12 GB of RAM. In addition, we estimate both forward and backward displacements in our unsupervised training framework, which further intensifies the memory limitation. 
	
	To mitigate the aforementioned problem, we employ gradient checkpointing \cite{chen2016training}, where all values of forward pass are not saved into memory. When back propagation requires the values of the forward pass, they are re computed. According to \cite{chen2016training}, it decreases memory usage up to 10 times with a computational overhead of only 20-30$\%$. Moreover, the network's weight are kept fixed in backward flow computation (the gray block in Fig. \ref*{fig:Unlight}) to further reduce the memory usage. 
	
	Envelope of RF data along with RF data and imaginary parts of analytic signal were used as three separate channels. As suggested by \cite{tehrani2020displacement}, envelope along with RF data is used to compensate the loss of information in RF data by the downsampling steps in the network. 
	
	We used the pre-trained weights of \cite{hui2018liteflownet}, which was obtained by training on 72,000 pairs of simulated computer vision images with known optical flows. Fine-tuning was performed on 2200 pairs of real ultrasound RF data with unknown displacement maps. The test images were not seen by the network during fine-tuning. The network was trained for 20 epochs on NVIDIA TITAN V using Adam optimizer. The learning rate was set to 4e-7 and the batch size was 1 due to memory limitations. Regarding the weights of each part of the loss function, there is a trade-off between bias and variance error. Higher weights of the smoothing loss result in smoother but more biased results and lower weights lead to lower bias but higher variance. We empirically set $\lambda_1$, $\lambda_2$ and $\lambda_3$ to 0.5, 0.005, 0.2, as we observed that these weights had a good balance between bias and variance error.
    
	\section{Results}
	We validated our proposed unsupervised fine-tuning using an experimental phantom and \textit{in vivo} data. We compare LiteFlowNet, our unsupervised fine-tuned LiteFlowNet and GLobal Ultrasound Elastography (GLUE) \cite{hashemi2017global}, which is a well known non-deep learning elastography method.  Codes associated with all of these methods are available online. 

	\subsection{Quantitative Metrics}
	Contrast to Noise Ratio (CNR) and Strain Ratio (SR) are two popular metrics used to assess the elastography algorithms in experimental phantoms and \textit{in vivo} data where the ground truth is unknown. These metrics are defined as \cite{ophir1999elastography}:

	\begin{equation}
	\label{Eq:SRCNR}
	SR =\frac{\overline{s}_{t}}{\overline{s}_{b}},\quad \quad CNR = \sqrt{\frac{2(\overline{s}_{b}-\overline{s}_{t})^{2}}{{\sigma _{b}}^{2}+{\sigma _{t}}^{2}}},
	\end{equation}	
	\noindent
	where $\overline{s}_{b}$ and $\overline{s}_{t}$ are average values of strain in the background and target regions of the tissue, and $\sigma _{b}$ and $\sigma _{t}$ denote variance values of strain in the background and target regions, respectively. CNR is a proper metric to measure a combination of bias and variance error. SR sheds light on the estimator bias in real experiments wherein the ground truth strain values are unknown \cite{tehrani2020displacement}. 
	\begin{figure*}[t]
		\centering
		\begin{subfigure}{0.24\linewidth}
			\centering
			\includegraphics[height=0.12\textheight]{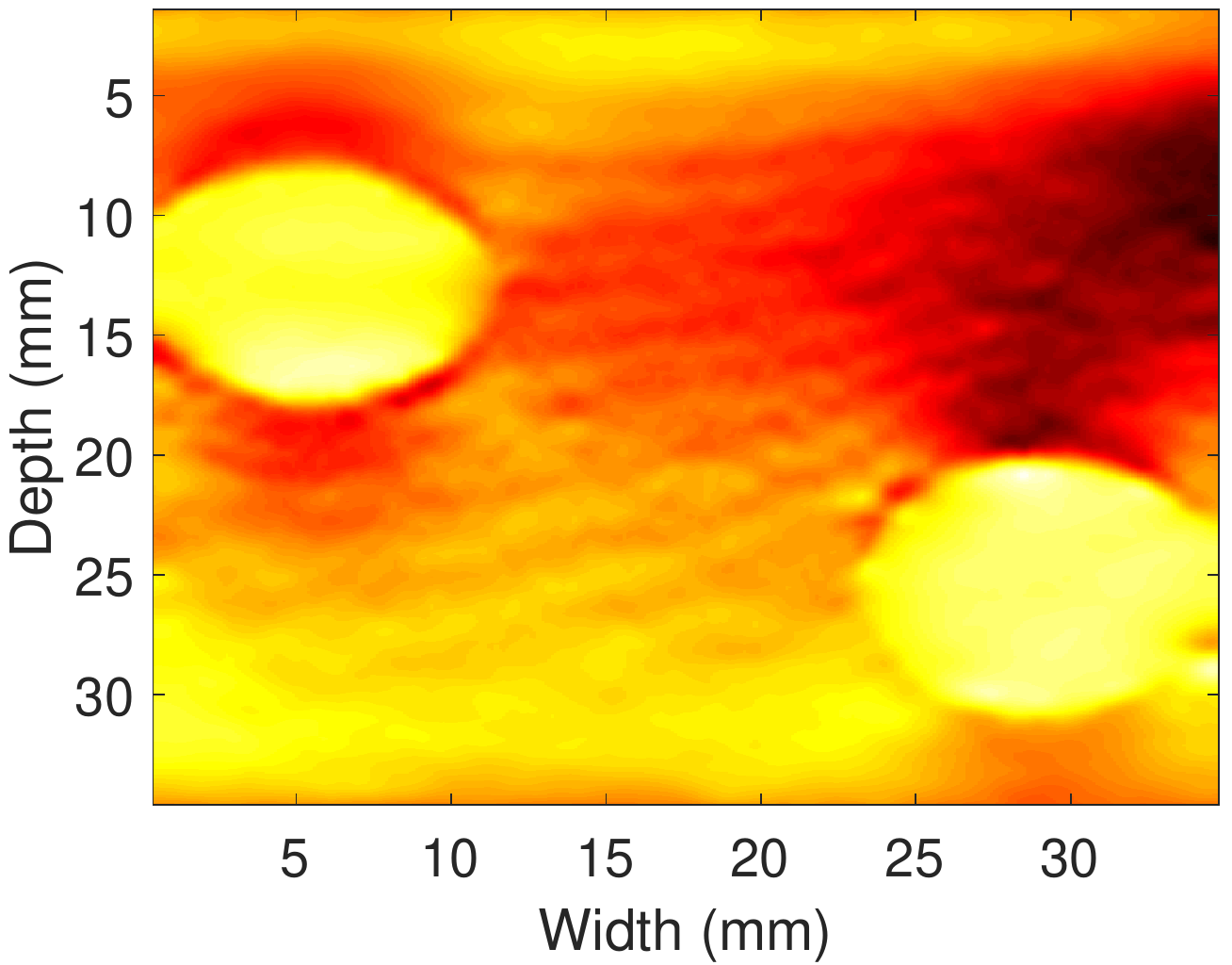}
			\caption{GLUE }\label{fig:image13}
		\end{subfigure}
		\begin{subfigure}{0.24\linewidth}
			\centering
			\includegraphics[height=0.12\textheight]{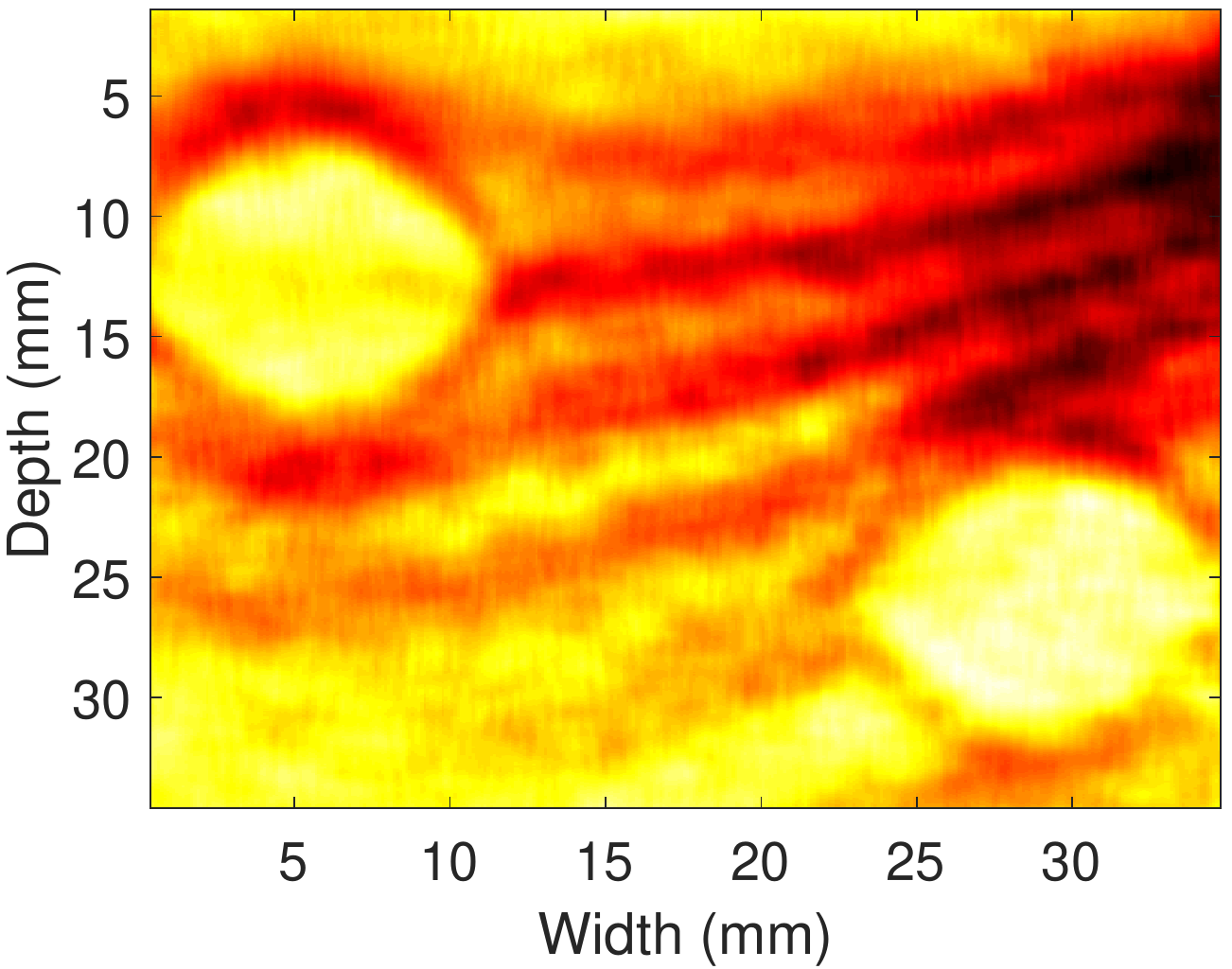}
			\caption{LiteFlowNet}\label{fig:image3}
		\end{subfigure} 
		\begin{subfigure}{0.24\linewidth}
			\centering
			\includegraphics[height=0.12\textheight]{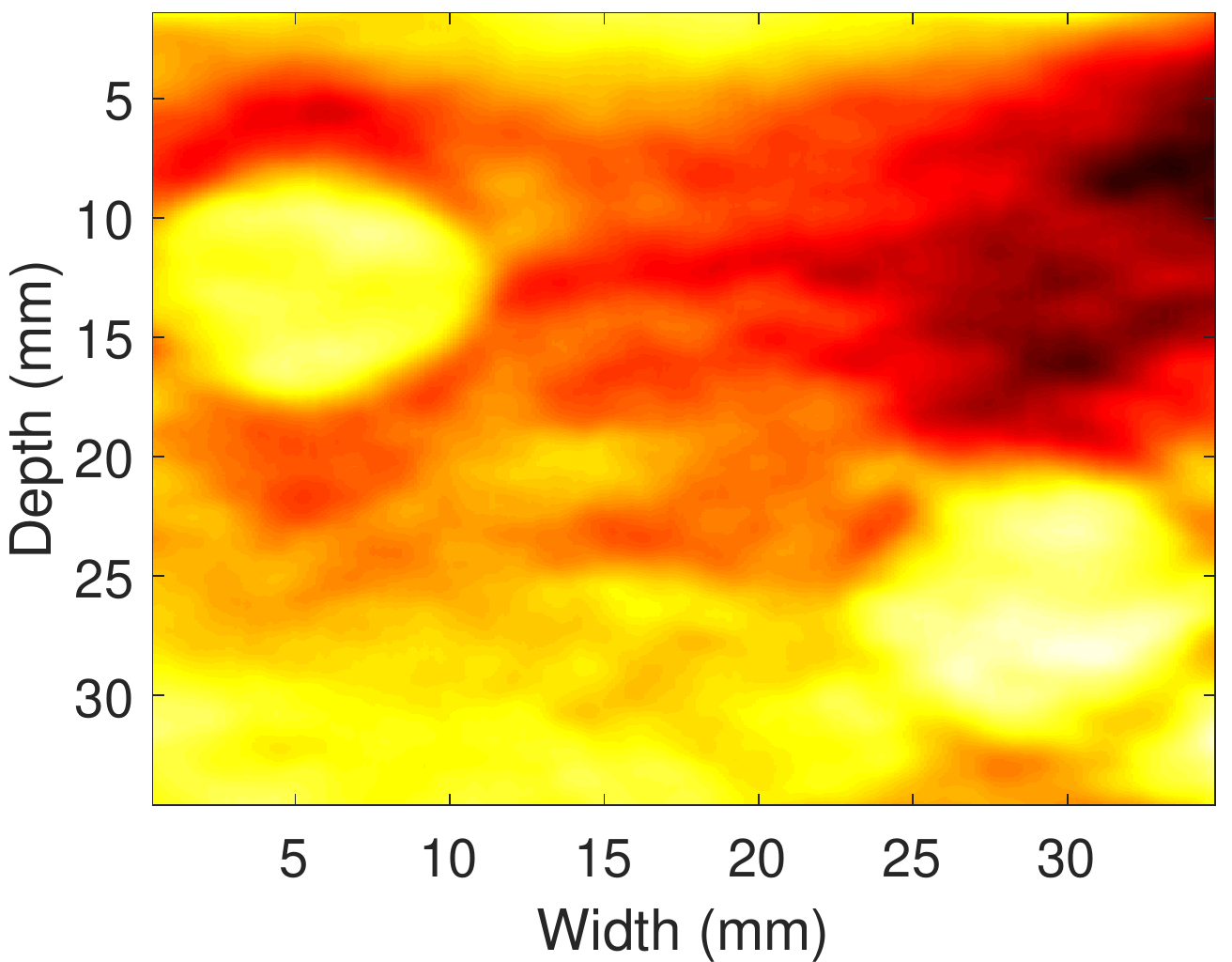}
			\caption{Unsupervised }\label{fig:image3}
		\end{subfigure}
		\begin{subfigure}{0.24\linewidth}
			\centering
			\includegraphics[height=0.12\textheight]{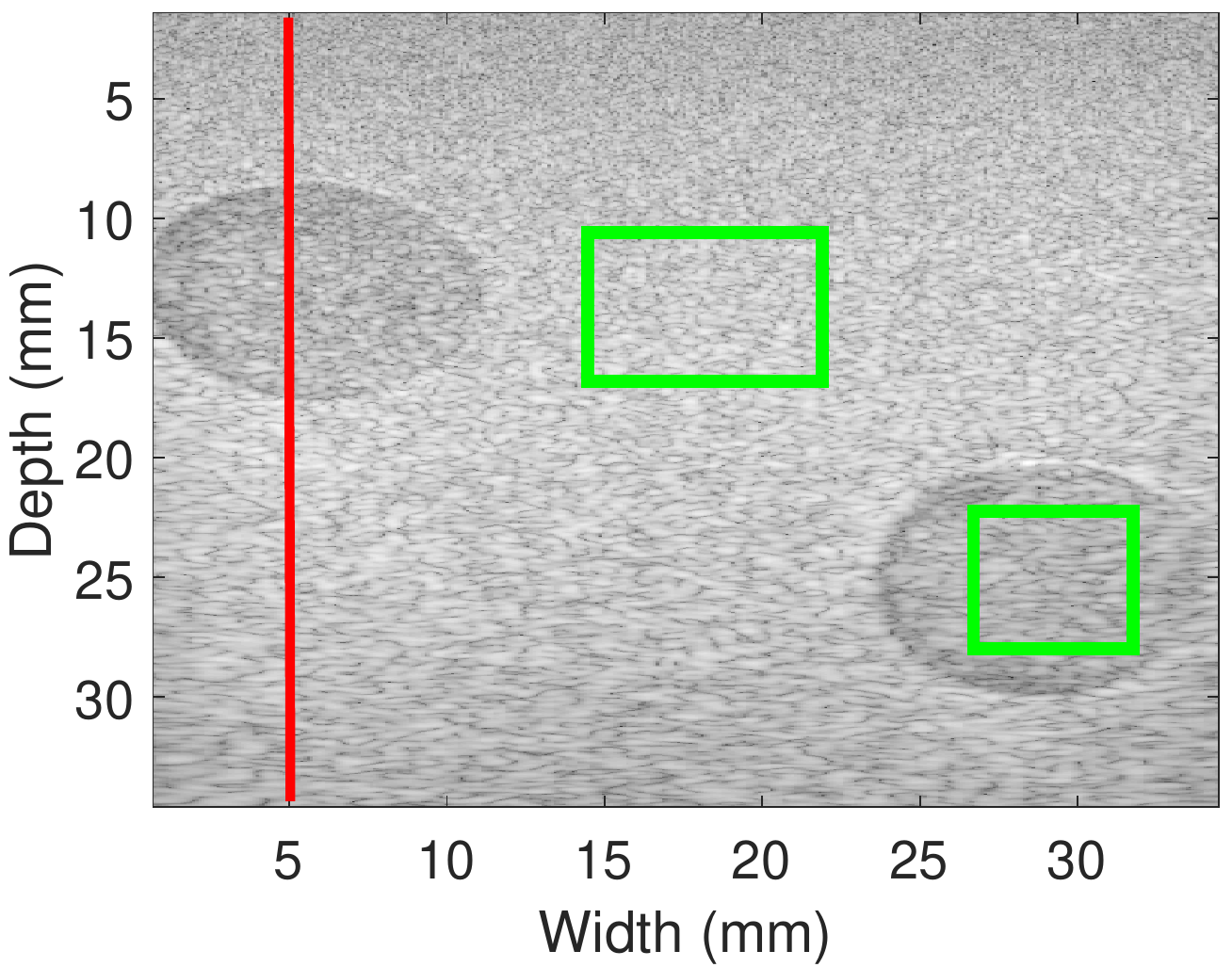}
			\caption{B-mode Image}\label{fig:image3}
		\end{subfigure}
		\begin{subfigure}{1\linewidth}
			\centering
			\includegraphics[height=0.035\textheight]{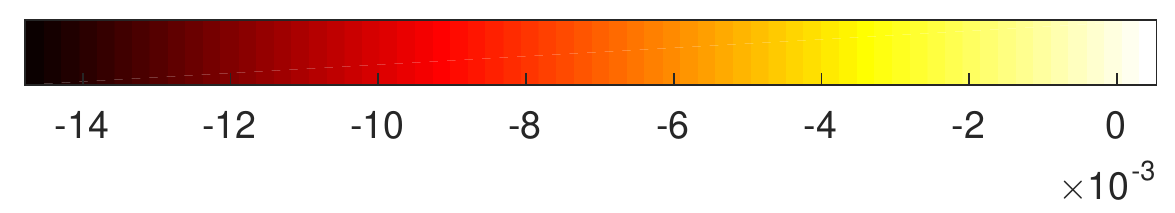}
		\end{subfigure}
		\RawCaption{\caption{Strain images of the experimental phantom with two hard inclusion. The windows used for CNR and SR computation are highlighted in the B-mode image (d). The strain of the line highlighted in red is shown in Fig. \ref{fig:line}.}
			\label{fig:phantom}}  	
	\end{figure*}

	\subsection{Experimental Phantom}

	We collected ultrasound images at Concordia University's PERFORM Centre using an Alpinion E-Cube R12 research ultrasound machine (Bothell,
	WA, USA) with a L3-12H linear array at the center frequency of 10 MHz and sampling frequency of 40 MHz. A tissue mimicking breast phantom made by Zerdine (Model 059, CIRS: Tissue Simulation \& Phantom Technology, Norfolk, VA) is used for data collection. The phantom contains several hard inclusions with elasticity values at least twice the elasticity of the tissue. The experimental phantom for test is from the same phantom but different part of the phantom is imaged. The composition of the phantom in test data is also different, where regions with only one inclusion are used for training and regions with two inclusions are used for testing. The test results are depicted in Fig. \ref{fig:phantom}. 
	
\begin{figure*}[t]
		\centering
		\begin{subfigure}{0.49\linewidth}
			\centering
			\includegraphics[height=0.16\textheight]{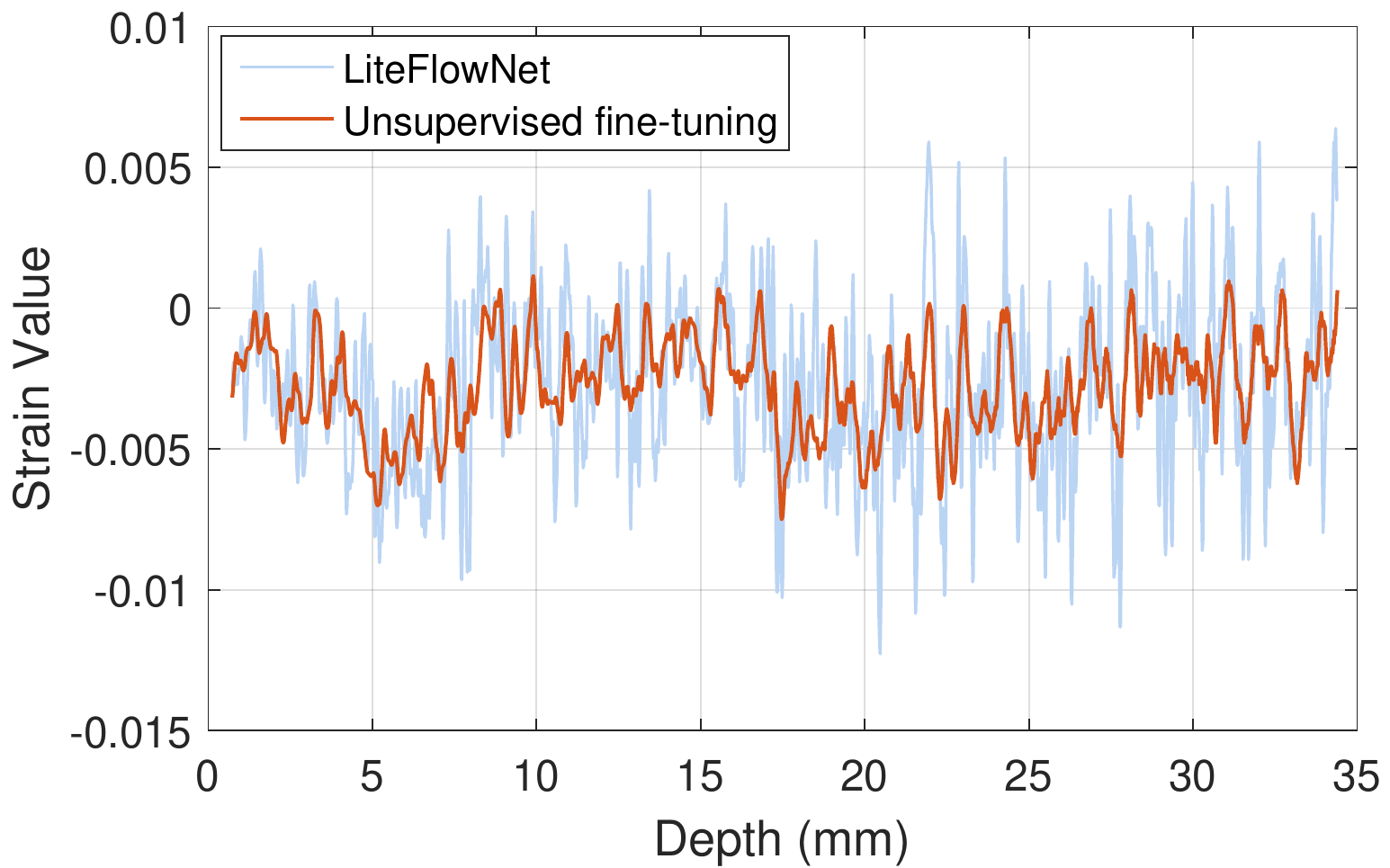}
			\caption{The strain of the line}
		\end{subfigure}
		\begin{subfigure}{0.49\linewidth}
			\centering
			\includegraphics[height=0.16\textheight]{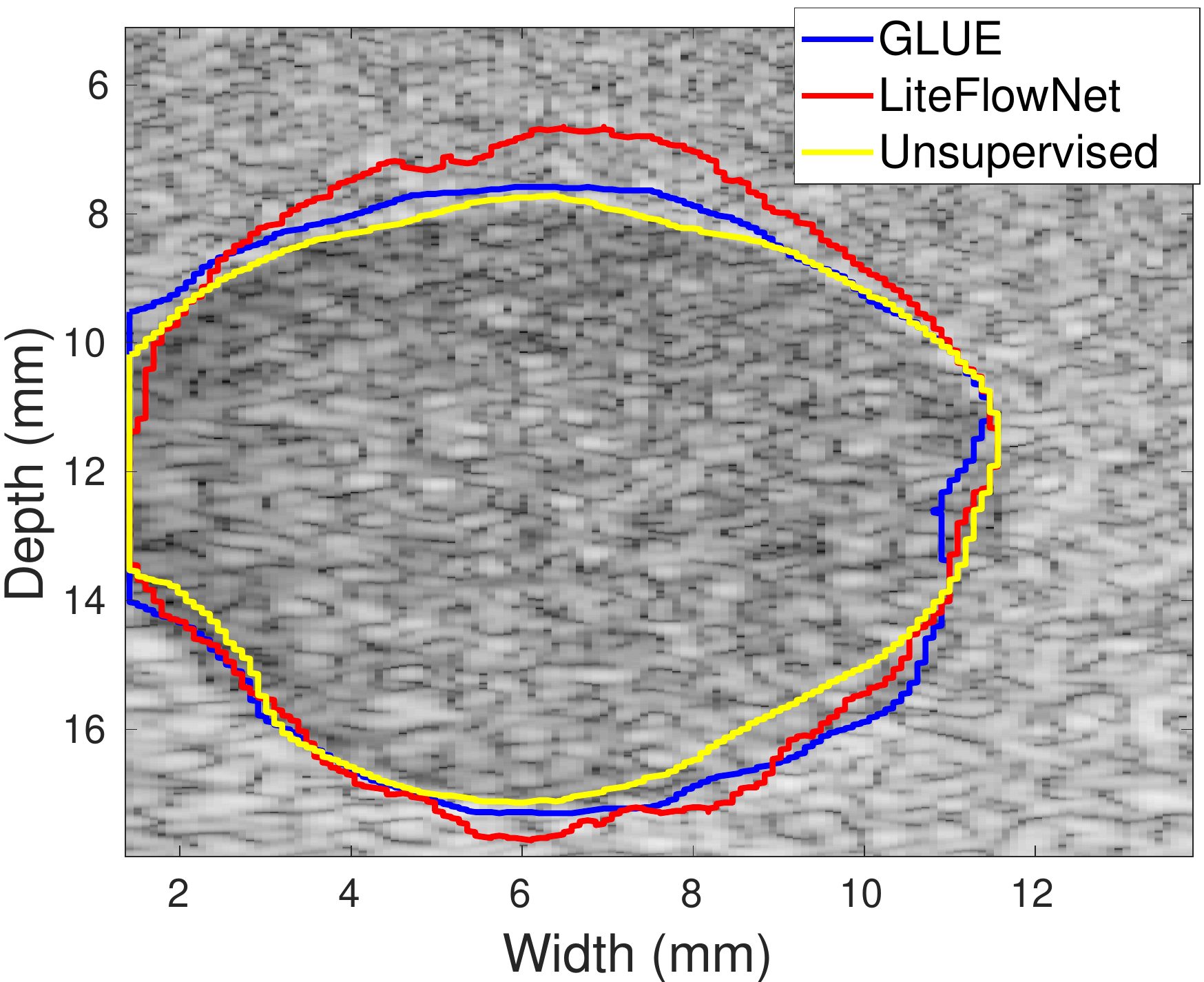}
			\caption{The Tumor at the top left}
		\end{subfigure} 
		\caption{The strain of the line specified in  Fig. \ref{fig:phantom} using small smoothing window (a). The top left tumor with the edges obtained by the three USE methods (b).}
		\label{fig:line}
	\end{figure*}

	The unsupervised fine-tuning improves the strain quality of LiteFlowNet producing a smoother strain with less artifacts. The quantitative results are shown in Fig. \ref{fig:cnr}. GLUE has the highest CNR which shows the high-quality strain but the SR is also the highest which indicates that it has the highest bias due to the strong regularization used in the algorithm. The unsupervised fine-tuning substantially improves the CNR of the network with very similar SR. In order to show the improvements better, the line specified in Fig. \ref{fig:phantom} (d) with small differentiation window is shown in Fig. \ref{fig:line} (a). The strain plots indicate that fine-tuning substantially reduces the variance error presented in the strain. We calculate the edges of the strain images using Canny edge detection and superimpose on the B-mode image to compare the size of different structures in in Fig. \ref{fig:line} (b). Here, the top left inclusion of Fig. \ref{fig:phantom} (d) is shown. It can be seen that LiteFlowNet substantially overestimates the size of the inclusion since the red curve is well outside of the inclusion. Our unsupervised fine-tuning technique corrects this overestimation.

	\subsection{\textit{In vivo} Data}  
\begin{figure}[t]
		\centering
		\begin{subfigure}{0.24\linewidth}
			\centering
			\includegraphics[height=0.12\textheight]{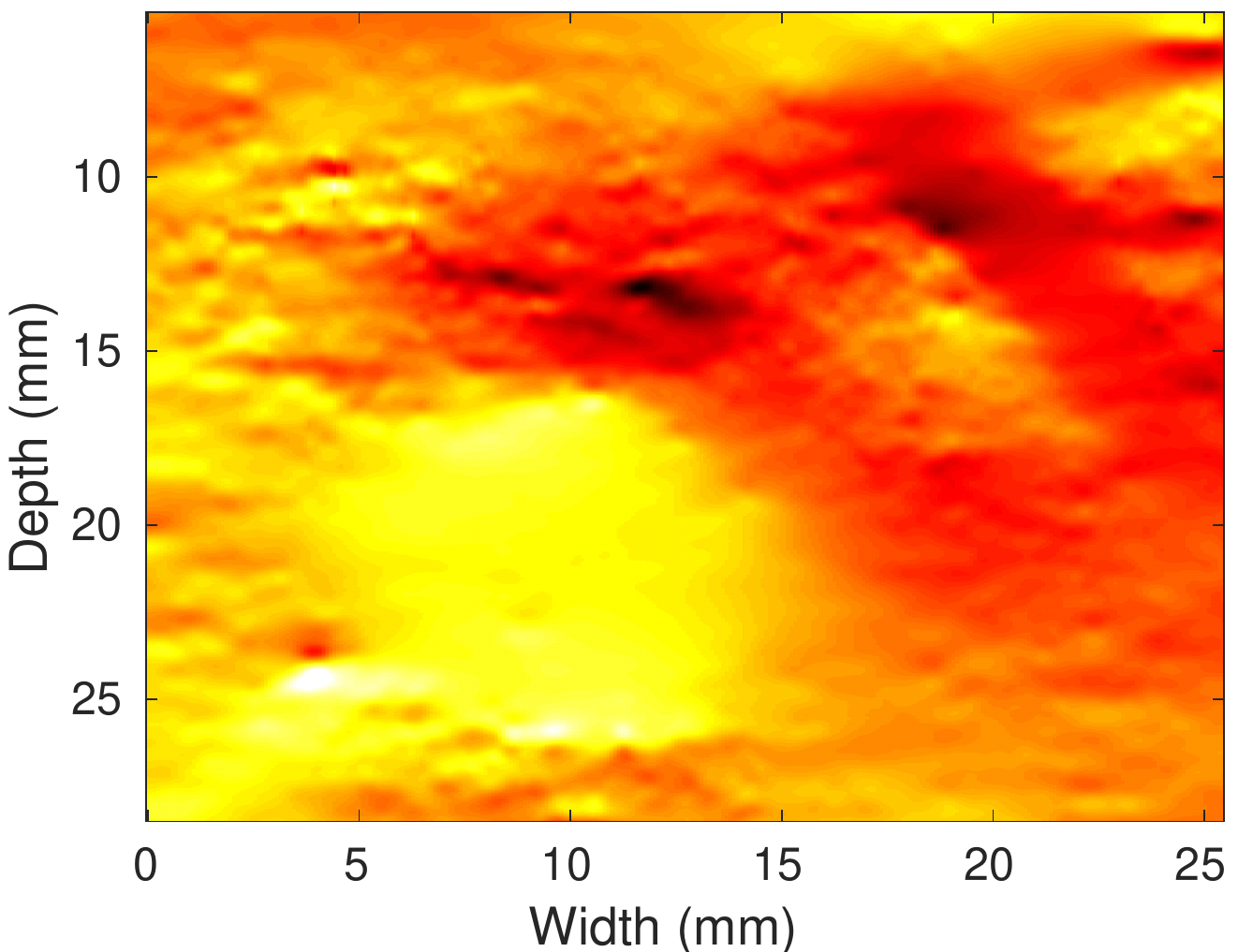}
			\caption{GLUE }
		\end{subfigure}
		\begin{subfigure}{0.24\linewidth}
			\centering
			\includegraphics[height=0.12\textheight]{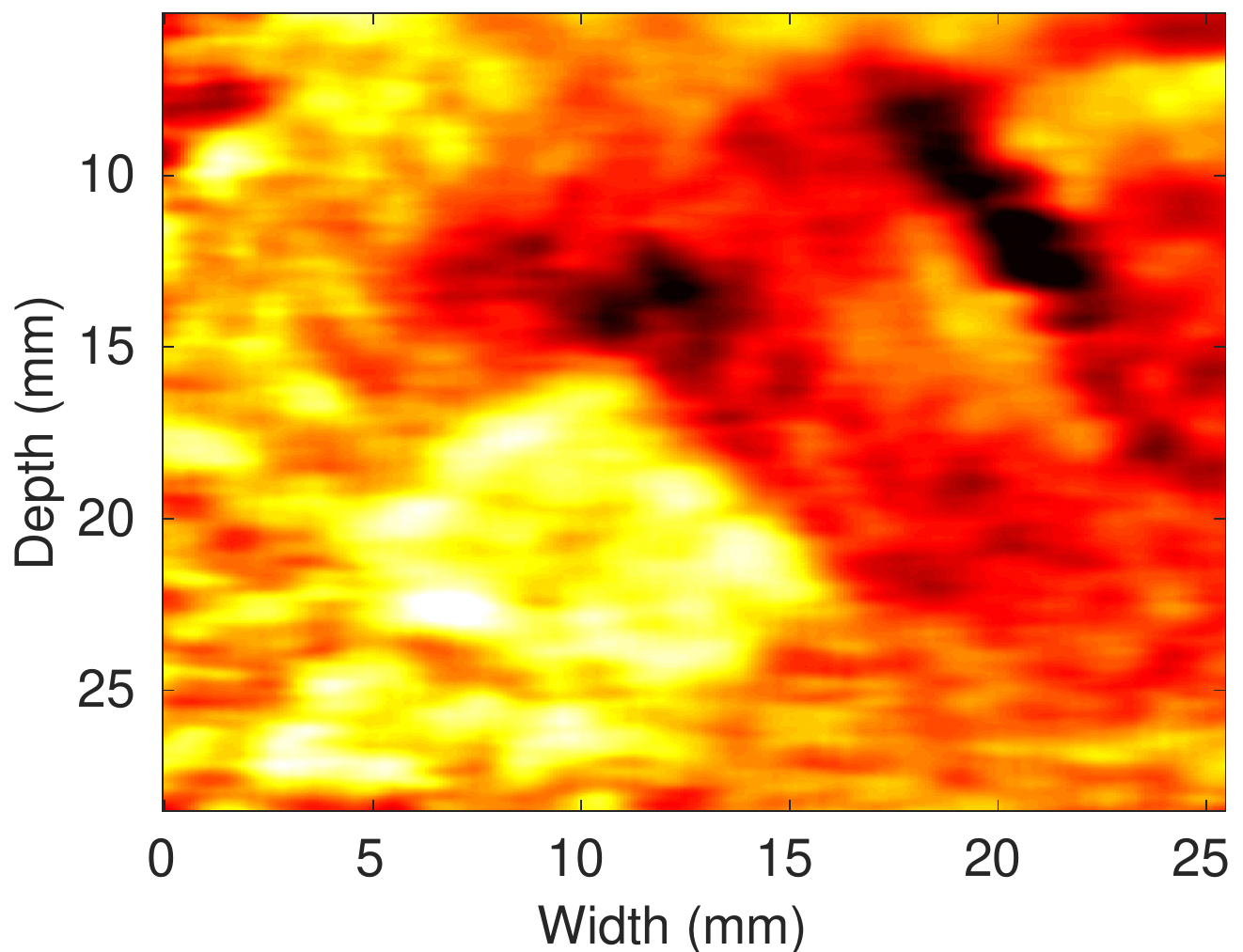}
			\caption{LiteFlowNet}
		\end{subfigure} 
		\begin{subfigure}{0.24\linewidth}
			\centering
			\includegraphics[height=0.12\textheight]{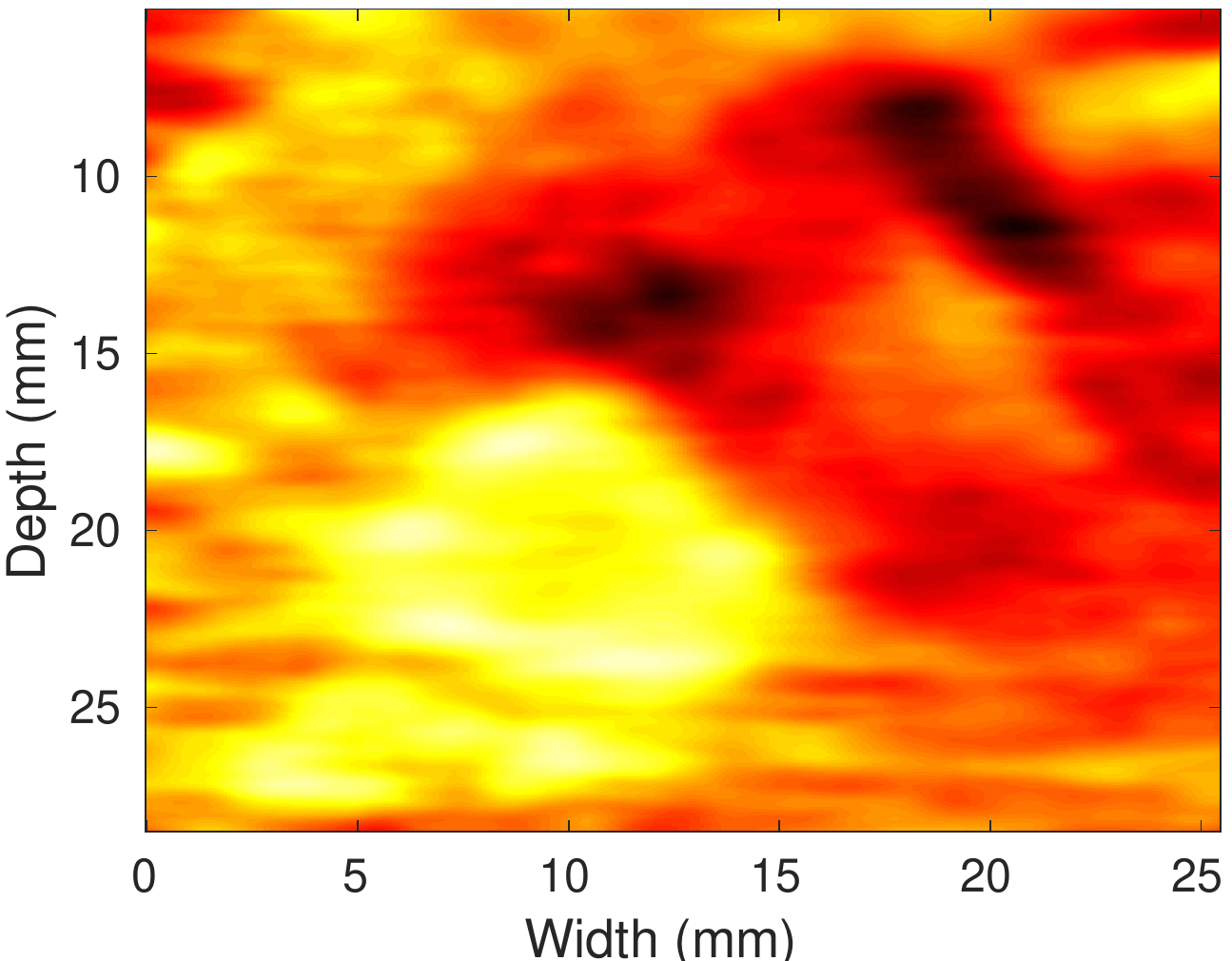}
			\caption{Unsupervised}
		\end{subfigure}
		\begin{subfigure}{0.24\linewidth}
			\centering
			\includegraphics[height=0.12\textheight]{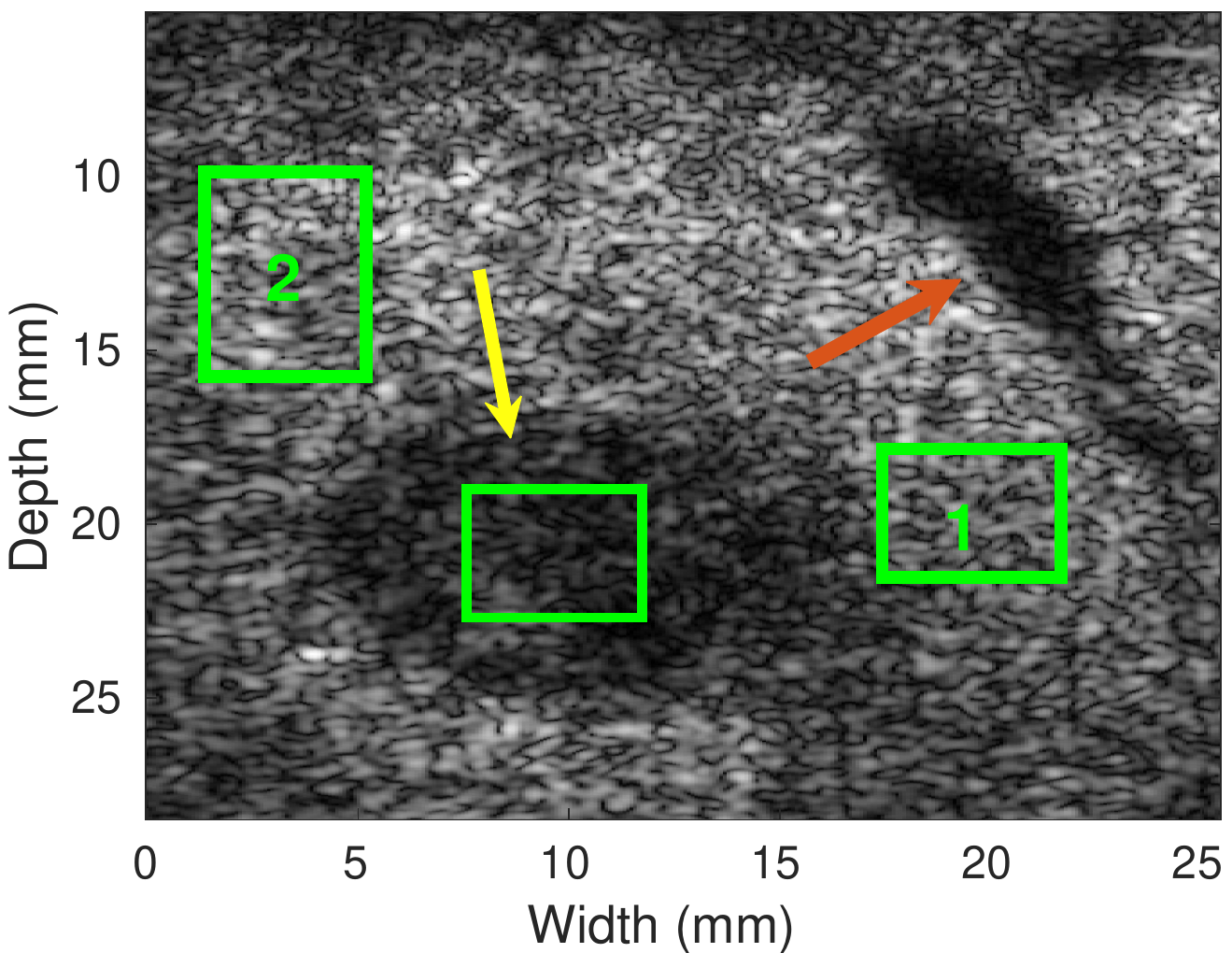}
			\caption{B-mode Image}
		\end{subfigure}	
		
		\begin{subfigure}{1\linewidth}
			\centering
			\includegraphics[height=0.035\textheight]{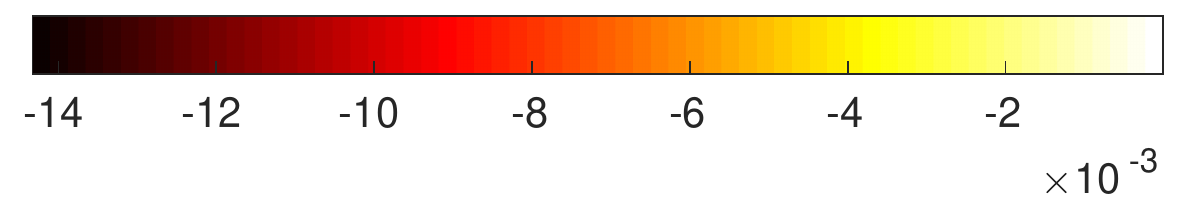}
		\end{subfigure}
		\RawCaption{\caption{Strain images of the \textit{in vivo} data. The two background and the target windows used for CNR and SR computation are highlighted by green in the B-mode image (d). The hard tumor and soft vein are marked by yellow and red arrows, respectively.}
		
			\label{fig:vivo}}  	
	\end{figure}

	\textit{In vivo} data was collected  at Johns Hopkins Hospital using a research Antares Siemens system by a VF 10-5 linear array with a sampling frequency of 40 MHz and the center frequency of 6.67 MHz. Data was obtained from patients in open-surgical RF thermal ablation for liver cancer \cite{rivaz2010real}. The study was approved by the institutional review board with consent of all patients. The strains obtained by the compared methods are shown in Fig. \ref{fig:vivo}. GLUE (a) produce high quality strain but it is over smoothed which is evident for the vein (it is marked by the orange arrow). LiteFlowNet (b) produces a strain with many artifacts and heterogeneities inside the tumor (the tumor is marked with the yellow arrow),  but it preserves the vein which is a small structure. Unsupervised fine-tuning (c) not only reduces the artifacts and heterogeneity inside the tumor, but also maintains the vein similar to LiteFlowNet.

\begin{figure*}[t]
		\centering
		\begin{subfigure}{0.48\linewidth}
			\centering
			\includegraphics[height=0.15\textheight]{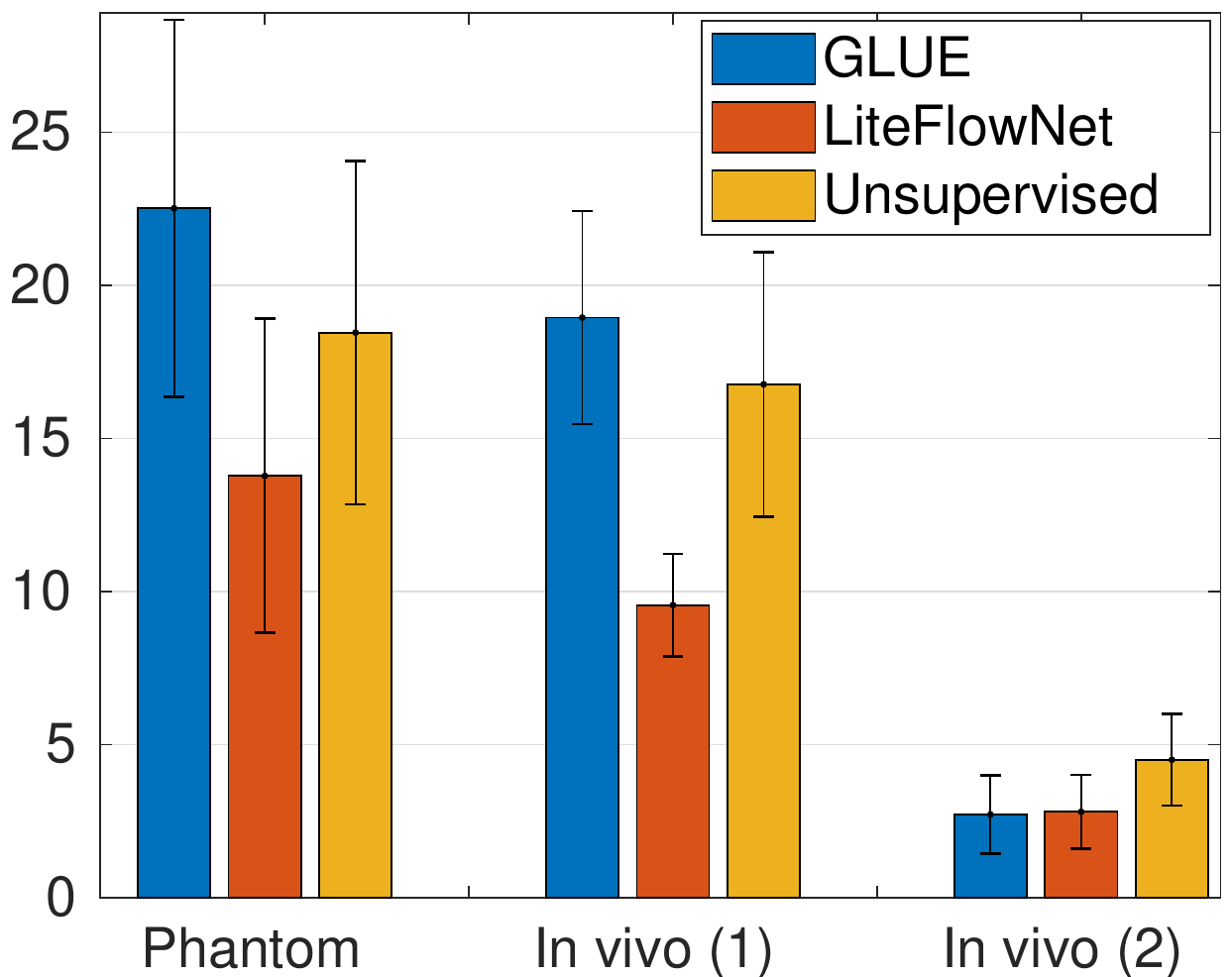}
			\caption{CNR (higher is better) }
		\end{subfigure}
		\begin{subfigure}{0.48\linewidth}
			\centering
			\includegraphics[height=0.15\textheight]{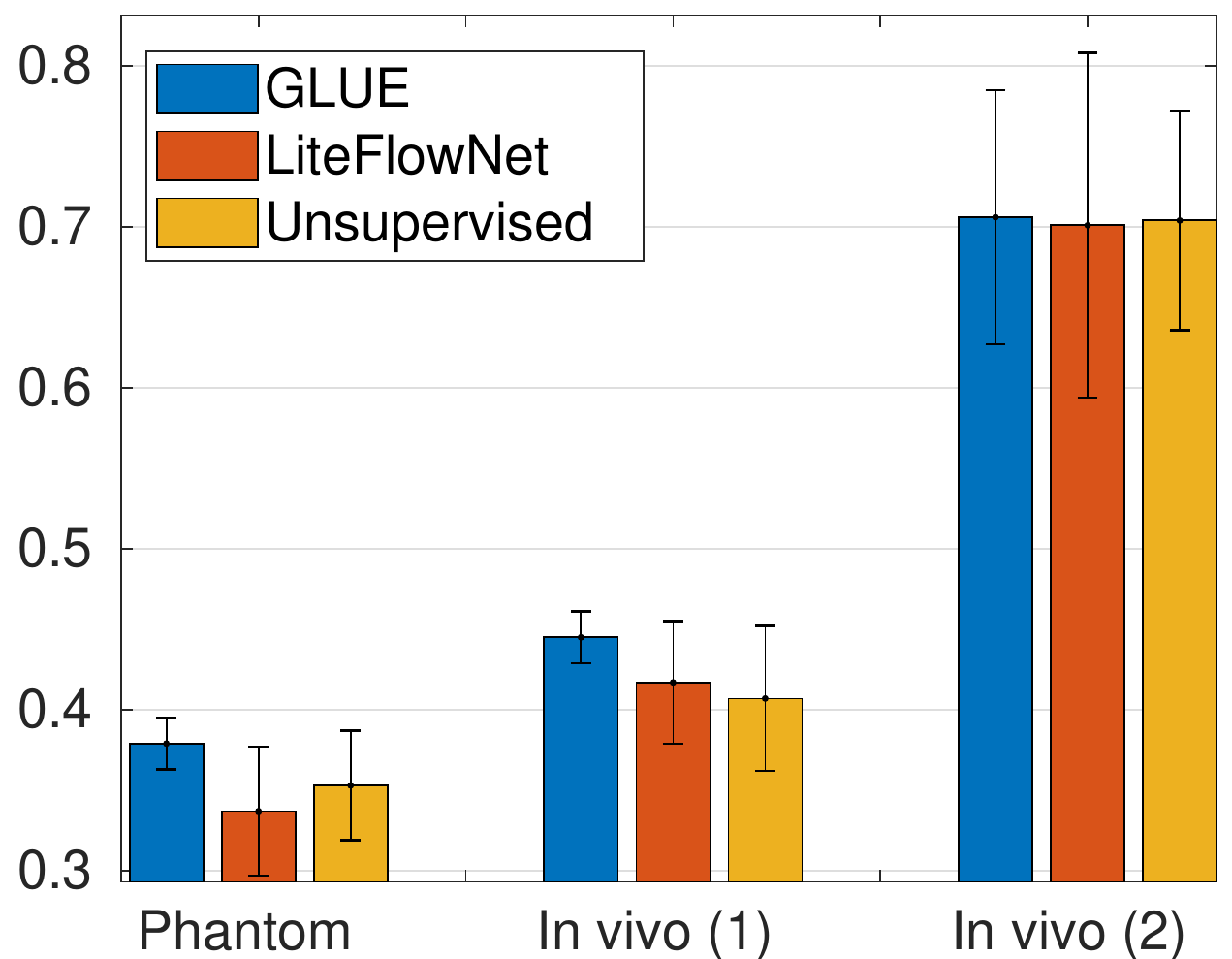}
			\caption{SR (lower is better)}
		\end{subfigure} 
		\RawCaption{\caption{CNR (a) and SR (b) of experimental phantom and \textit{in vivo} data. \textit{In vivo} (1) and \textit{in vivo} (2) correspond to the CNR with background of 1 and 2 in Fig. \ref{fig:vivo} (d), respectively.}
		
			\label{fig:cnr}}  	
	\end{figure*}

	The quantitative results are given in Fig. \ref{fig:cnr}. The first background window (1 in Fig. \ref{fig:vivo} (d)) is very different than the tumor region, GLUE and the fine-tuned network produce similar CNRs between the tumor and this window. The second background (2 in Fig. \ref{fig:vivo} (d)) has very high amount of artifacts and the strain value is very close to the strain of tumor. Fine-tuned network has the best CNR for this challenging background which can be confirmed by visual assessment of Fig. \ref{fig:vivo}. Regarding SR, LiteFlowNet has the lowest SR but the differences with the unsupervised fine-tuned network are negligible. Among the compared methods, GLUE has the worst SR results which indicates the high bias error presented in this method. 

	\section{Conclusion}
	Herein, we proposed an semi-supervised technique for USE. We fine-tuned an optical flow network trained on computer vision images using unsupervised training. We designed a loss function suitable for our task and substantially improved the strain quality by fine-tuning the network on real ultrasound images. The proposed method can facilitate commercial adoption of USE by allowing a convenient unsupervised training technique for imaging different organs using different hardware and beamforming techniques.  Inference is also very fast, facilitating the use of USE in image-guided interventions.
	\section{Acknowledgement}
	We thank NVIDIA for the donation of the GPU. The \textit{in vivo} data
	was collected at Johns Hopkins Hospital. We thank E. Boctor, M. Choti
	and G. Hager for giving us access to this data.
	\bibliography{paper2365}
	\bibliographystyle{splncs04}
	
	%
	%
	%
	%
	%
\end{document}